# Single Flux Quantum Based Ultrahigh Speed Spiking Neuromorphic Processor Architecture

Ali Bozbey[1,*], **Mustafa Altay Karamuftuoglu**[1], Sasan Razmkhah[1], Murat Ozbayoglu[2]

[1]Department of Electrical and Electronics Engineering, TOBB Economy and Technology University, 06560, Ankara, Turkey

[2]Department of Computer Engineering, TOBB Economy and Technology University, 06560, Ankara, Turkey
*bozbey@etu.edu.tr

**ABSTRACT**
Artificial neural networks inspired by brain operations can improve the possibilities of solving complex problems more efficiently. Today's computing hardware, on the other hand, is mainly based on von Neumann architecture and CMOS technology, which is inefficient at implementing neural networks. For the first time, we propose an ultrahigh speed, spiking neuromorphic processor architecture built upon single flux quantum (SFQ) based artificial neurons (JJ-Neuron). Proposed architecture has the potential to provide higher performance and power efficiency over the state of the art including CMOS, memristors and nanophotonics devices. JJ-Neuron has the ultrafast spiking capability, trainability with commodity design software even after fabrication and compatibility with commercial CMOS and SFQ foundry services. We experimentally demonstrate the soma part of the JJ-Neuron for various activation functions together with peripheral SFQ logic gates. Then, the neural network is trained for the IRIS dataset and we have shown 100% match with the results of the offline training with $1.2 \times 10^{10}$ synaptic operations per second (SOPS) and $8.57 \times 10^{11}$ SOPS/W performance and power efficiency, respectively. In addition, scalability for $10^{18}$ SOPS and $10^{17}$ SOPS/W is shown which is at least five orders of magnitude more efficient than the state-of-the-art CMOS circuits and one order of magnitude more efficient than estimations of nanophotonics-based architectures.

## Introduction

The scientific community enthralled by understanding the general principles of human brain functions, as a further matter, on how to mimic the abilities by utilizing artificial neurons for more efficient computing. So far, almost all the general computing systems are based on Von Neumann architecture and due to Von-Neuman bottleneck[1], development of alternative devices and architectures are inevitable. Additionally, quantum computers are recently emerging and quantum supremacy is approaching[2]. We believe that soon, all the positive aspects of all three computation architectures will be hybridized. It is known that Von Neumann architectures and neuromorphic architectures outperform each other in the execution of different type of tasks. Additionally, neuromorphic computers offer a different way of computation and a robust way of emulation for Artificial Intelligence (AI) implementations than conventional computer structures. The capabilities of these computers overcome the known challenges such as dynamic learning ability, dynamic network remapping and energy efficient computation for networks. Generally, neuromorphic processors are often implemented using the VLSI-based CMOS with little resemblance to the their biological counterparts and the challenge is to exploit a wider range of biological principles at the hardware or the model such as neuro-mimicking materials and principles [3–6], spiking [7–9], rate-based and population-level neuronal dynamics [10–12] and plasticity [13,14].

The neuron is considered as a fundamental unit of human brain due to the functions, receiving and sending electro-chemical signals to process the data and creating overall behavior[15]. Dendrites receive synaptic inputs from other neuron axons, and they bring information to the cell body. Soma collects all signals from its dendrites and creates a relative response that depends on received signals. The axon carries an electrical response to the connected neurons. The functionality comes from self-assembly of brain cells, known as nerve cells or neurons and mathematical neuron models are created with relatively similar components of biological structure[16,17]. Additionally, Artificial Neural Network (ANN) is considered as the alternative and effective way to deal with complex problems such as image recognition, decision making, and forecasting while simulating the biological brain[18,19]. Implementation of the neuron behavior gives an opportunity to create neuromorphic computers with the ability of learning events just like the way brain does. Computational software tools connect artificial neurons to each other to create ANN to adopt the biological neural network behavior. ANN software tools have gained extensive acceptance for neural network applications because of the learning abilities, and computational power and speed through parallel processing. Even though there are hardware neuron design examples based on CMOS devices[20–23], CMOS technology is facing its fundamental limits as Moore's law[24–26] reaches its end, and this motivates the different technology investigations about artificial neuron applications[27,28] for implementing a neuromorphic computer.

Single flux quantum (SFQ) technology[29,30] is one of the strongest candidates for neuromorphic processor implementation technologies. Characteristic features of Josephson junctions (JJs) which have ultra-high-speed switching behavior with low power consumption match the properties of biological neurons. The comparison of biological neuron[15] with CMOS Integrate and Fire Neuron (IFN) Model representations[20–23,31] and our superconducting IFN Model features are shown in Supplementary Information Table S *1*. As shown, an SFQ neuron is six orders of magnitude faster and 3 orders of magnitude energy efficient. This is enabled by the possibility to implement Josephson junction-based soma, synapse, and axon structures which are inherently based on energy efficient and ultra-fast spiking topologies. Furthermore, ability to transmit data to long distances with the speed of the light without any power consumption by using passive transmission lines[32] removes the limitations on the practical chip dimensions and opens the possibilities of high-speed multi-chip module configurations[32,33].

There are several research papers that show the implementation of brain cell characteristics by using SFQ technology[34–49] and the related comparison between designs is given in the Supplementary Information Table S *2*. As shown, neurons based on Josephson junctions have been described in the literature beginning in early 90s. However, all of them have remained at the level of simulation and/or single neuron level. None of them have proven to be scalable, practically implementable, and trainable over the course of three decades. Our proposed design is based on leaky IFN model and it demonstrates effective and robust way of the implementation of a biological brain cell operation. It has all the required characteristic features such as trainability with commodity software even after fabrication, compatibility with standard SFQ logic gates, high computation power and power efficiency. Another main advantage of the designed neuron is its compatibility with the established SFQ and CMOS foundry services[50–53]. For the first time, we have experimentally shown that an SFQ based JJ-Neuron can be implemented, reliably operated and be used to design an artificial neural network with established neural network training methods and tools. For this purpose, we chose a publicly available benchmarking dataset and implemented it using the proposed neurons with 100% match with the results of the offline training. Then we discuss the scalability potential of the JJ-Neuron based ANNs and compare it with the alternative technologies in terms of performance and efficiency to show that is among the highest potential ones.

## Results

For a biological neural network, synaptic strengths define the function of the neural network, and the network provides the same operation if synapses remain unchanged. If the synapse values are steady, biological memory function of network will be achieved[54]. If a person does the action repetitively, related synaptic weight values increase, and this provides recall function of the network. In a neuron structure, main body of the neuron, soma, acts like a temporary storage of inputs to be able to do aggregation function. If enough input pulses arrive to soma from the synapses, it fires an axonal (output) pulse and the operation is called somatic operation. However, if the number of inputs is not enough within a certain time, no output pulse will be released. These principles are inherently implemented in JJ based logic circuits, namely SFQ circuits[29,30] where the logic "1" and "0" is based on SFQ pulses. Thus, it is possible to implement the operation of a biological neuron by JJs and SFQ circuits in an intuitive way.

In this research, our circuit implements the mathematical model named McCulloch-Pitts neuron[55] which is mainly composed of a summation function and an activation function. In the model, number of inputs and their individual weighs determine the function of the cell. Operation starts with individual multiplications of inputs ($x_k$) with their weights ($w_k$) as shown in Eq. (1). Input weights represent biological neuron synapses and it simulates chemical transmission of the neuron connection. Negative values of the weight have inhibitive effects while positive values of the weight are considered as actuators. Then, all values arriving from the inputs are added together on the summation function. If the sum of multiplication results ($u$) surpasses the threshold value ($\theta$), the activation function provides a single output and the pulses will be transmitted to the following neurons via synapses as reported in Eq. (2). The combination of summation and activation functions creates the functional cell body of a single neuron and the activation function is a control component that adjusts the amplitude of neuron output ($Y$). Additionally, the mathematical model representation of an artificial neuron is given in the Supplementary Information Figure S 1.

$$u = \sum_{k=1}^{N} x_k w_k \qquad (1)$$

$$Y = f(u) = 1 \ \ if \ u \geq \theta$$
$$Y = f(u) = 0 \ \ else \qquad (2)$$

We designed an SFQ based ultra-fast artificial neuron that can implement all the characteristic features of an artificial neural network including the ability to learn and recall. To test the applicability of the proposed design, we chose the IRIS dataset from UCI repository[56] which consists of 150 data points for 3 different Iris plant types. In the neural network, we had 4 inputs, 4 hidden neurons and 3 output neurons. The combination of neuron blocks creates a fully connected feed-forward type neural network. The details of the neural network development which includes backpropagation with gradient descent training and the discrete weight determination through genetic algorithm optimization are provided in detail in the Supplementary Information.

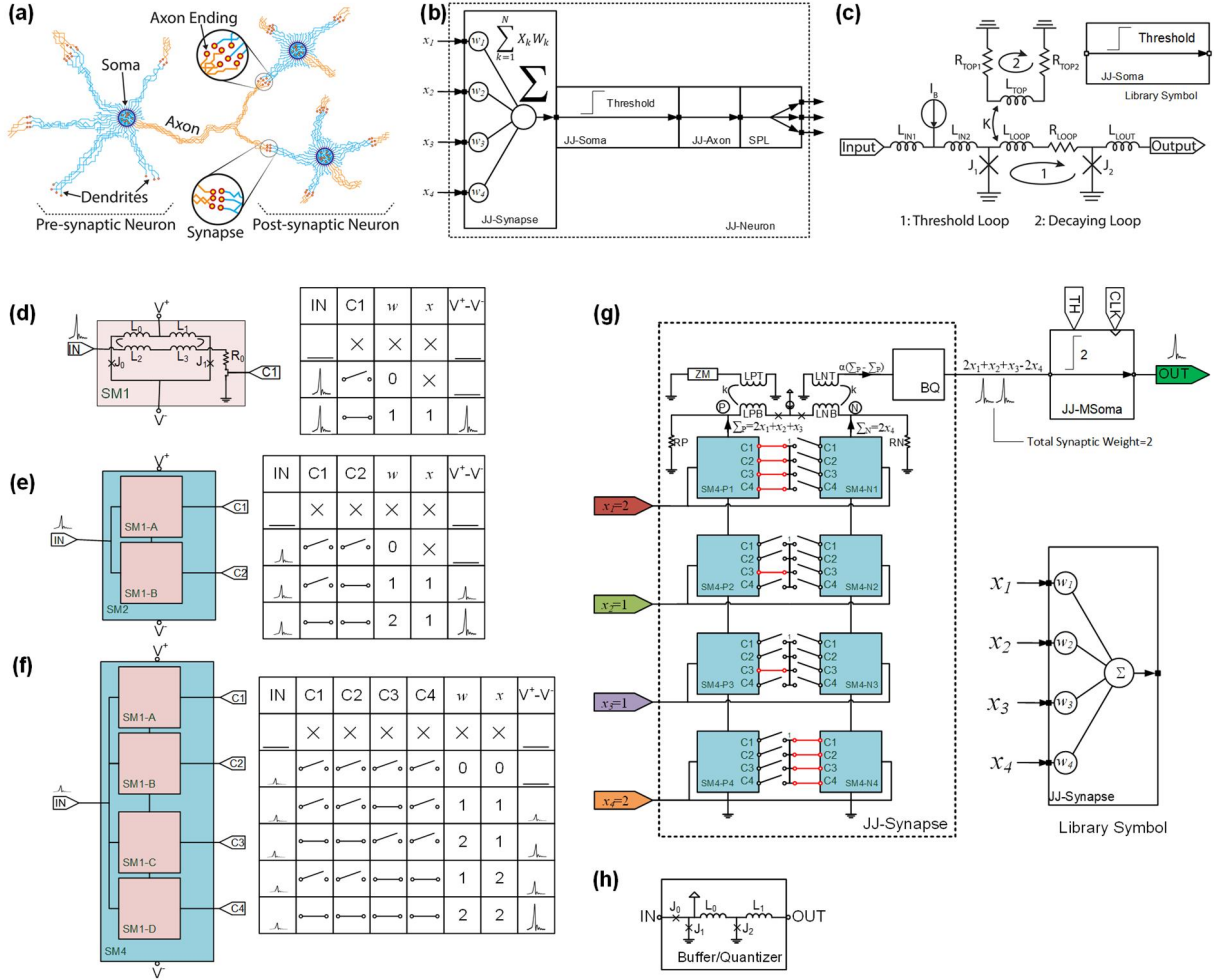

**Figure 1** The components of typical biological neuron cell (**a**) and an SFQ based spiking neuron cell (JJ-Neuron) (**b**). Biological (artificial) neuron consists of three parts: Synapse (JJ-Synapse), soma (JJ-Soma), and axon (JJ-Axon). Multiple connections between JJ-Soma output and the following neuron dendrite are achieved with Pulse Splitter (SPL) circuit. The JJ-Soma cell (**c**) contains two main loops that are mutually connected to each other. The threshold loop adjusts the current's limit for excitatory inputs and the decaying loop helps to change the threshold loop's time constant with the mutual coupling value for the current dissipation. Serially connected SQUID loops (**d**), (**e**), (**f**) and a Buffer/Quantizer (BQ) circuit (**g**) form the JJ-Synapse circuit (**h**). Unit cell of the synapse is SM1. When C1 switch is closed and an SFQ pulse is applied at the input IN, current flowing on the L2 and L3 is coupled to the SQUID loop J0, L0, J1, L1. So, an SFQ voltage is observed between nodes $V^+$ and $V^-$. This case corresponds to weight ($w$)=1, and $x$=1. When C1 switch is open, no current flows from L2 and L3. In this case, $V^+ - V^- = 0$ since no current is coupled to the SQUID loop ($w$=0). Details of the SM2, SM4 and implementation of negative weights are presented in Supplementary Text. × in the tables denote the "don't care" condition. That is, if IN=0, regardless of Ci, $w$, or $x$, output=0. For the SM1 circuit, $J_0=J_1=109.0\mu A$, $L_0=L_1=30.12pH$, $L_2=L_3=30.12pH$, $R_0=1.02\Omega$, LPT=LNT=8.00pH, LPB=LNB=5.80pH, RP=RN=0.98Ω. For the BQ circuit, $J_0=133~\mu A$, $J_1=112~\mu A$, $J_2=189~\mu A$, $L_0=3.91pH$, $L_1=1.323pH$.

## Artificial Neuron Implementation

Proposed artificial neuron consists of artificial synapse, soma, axon, and dendrite circuits as shown in Figure 1(a, b). Spatial and temporal integration of SFQ pulses are realized by using the Josephson junction (JJ) based artificial synapse circuit (JJ-Synapse). Somatic operation is carried out by JJ based artificial soma (JJ-Soma) circuit. Axon structure is created with Josephson Transmission Line (JTL) circuits (Figure S 2) and multiple connections between JJ-Soma output and the following neuron dendrite are achieved with Pulse Splitter (SPL) circuits (Figure S 3).

JJ based artificial spiking neuron and biological neuron are analogous to each other and it is possible to implement most of the functionalities of biological neuron with the presented circuit. However, this property alone is not enough for practical applications. For implementing a complicated artificial neural network, many of these cells should be able to reliably fabricated, and neuron circuits should able to be interfaced with conventional logic circuits as well as the other neuron circuits for input and output signals. As the conventional logic interface, we aimed to match the JJ-Neuron's parameters with the SFQ logic circuits since the SFQ logic technology is already matured for the implementations of high speed and complicated logic circuits[57–64]. In addition, SFQ circuits and proposed artificial soma cell body can be fabricated by using the same foundry process on the same chip. So, cost and reliability of fabrication and ability to use available design tools enable convenient scaling of the artificial neural network with the proposed artificial neuron. A JJ-Neuron may have many inputs and outputs depending on the network topology, each of which is compatible with the following neurons and SFQ logic circuits[29,65,66]. During the optimization, the JJ-Neuron is connected to JTL cells on input and output lines since the impedances of all the cells in the library are the same as a JTL. To optimize the circuits, we selected Particle Swarm Optimization (PSO) because it is one of the non-linear optimization methods and its algorithm is based on mimicking the swarm particles. Each particle that seeks for the maxima point in each search-space determines the values based on the objective function of the application. Further information can be found in [57,67] and in the optimization section of Supplementary Information.

## Synapse

One of the most critical parts of implementing an artificial neural network is to have synapses with programmable weights and ability to handle various input levels while keeping the circuit size, speed, and power consumption at reasonable values. In addition, the synapses should be able to handle both the excitatory and inhibitory inputs. We designed a synapse circuit (JJ-Synapse) that consists of serially connected SQUID loops and input LR circuits, each of which is coupled with a SQUID loop and matched with the outputs of the SFQ library elements as shown in Figure 1(d-h). Since the building blocks of the synapse circuit (SM1) are serially connected, total bias current and power consumption is substantially reduced. Each synapse circuit accepts spiking inputs and generates the positive or negative weighted output for the next layer. Principle of operation of the JJ-Synapse is explained in the Supplementary Text.

## Soma

To imitate the somatic operation and build artificial soma structure, we modelled the neuron cell body by using Josephson junctions and passive elements, and the schematic of JJ-Soma is shown in Figure 1(c). After the synaptic weight reaches the threshold value of JJ-Soma, a single SFQ pulse will be generated at the neuron output. The interconnection between neurons is done by SFQ circuits that enable to build complex structures.

The circuit is mainly formed by a threshold loop, a decaying loop, and a mutual inductance between the threshold loop and the decaying loop. The loops adjust the fading time of pulses that are held in the threshold loop. Various combinations of parameter values can create different threshold values and decaying times on the neuron circuit.

Furthermore, it is possible to implement an multiple threshold JJ-Soma (JJ-MSoma) by designing JJ-Soma circuits with various threshold values as shown in Figure S *6*. Then by enabling the bias switches T1-T5 or DFFe clock inputs CLK1-CLK5, based on the training dataset, desired threshold parameters can be achieved even after fabrication that enables training of the network. Especially in the case of using the bias switch option, there is no power consumption overhead since the unused JJ-Soma circuits are not biased.

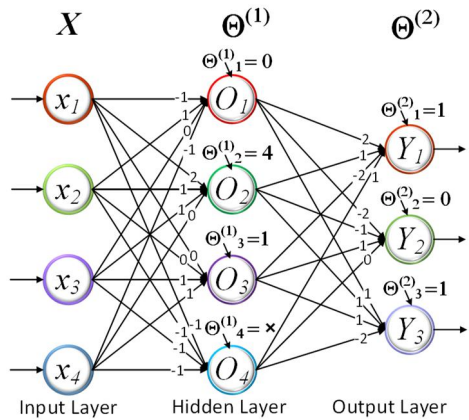

Figure 2 Topology of the neural network for the Iris dataset.

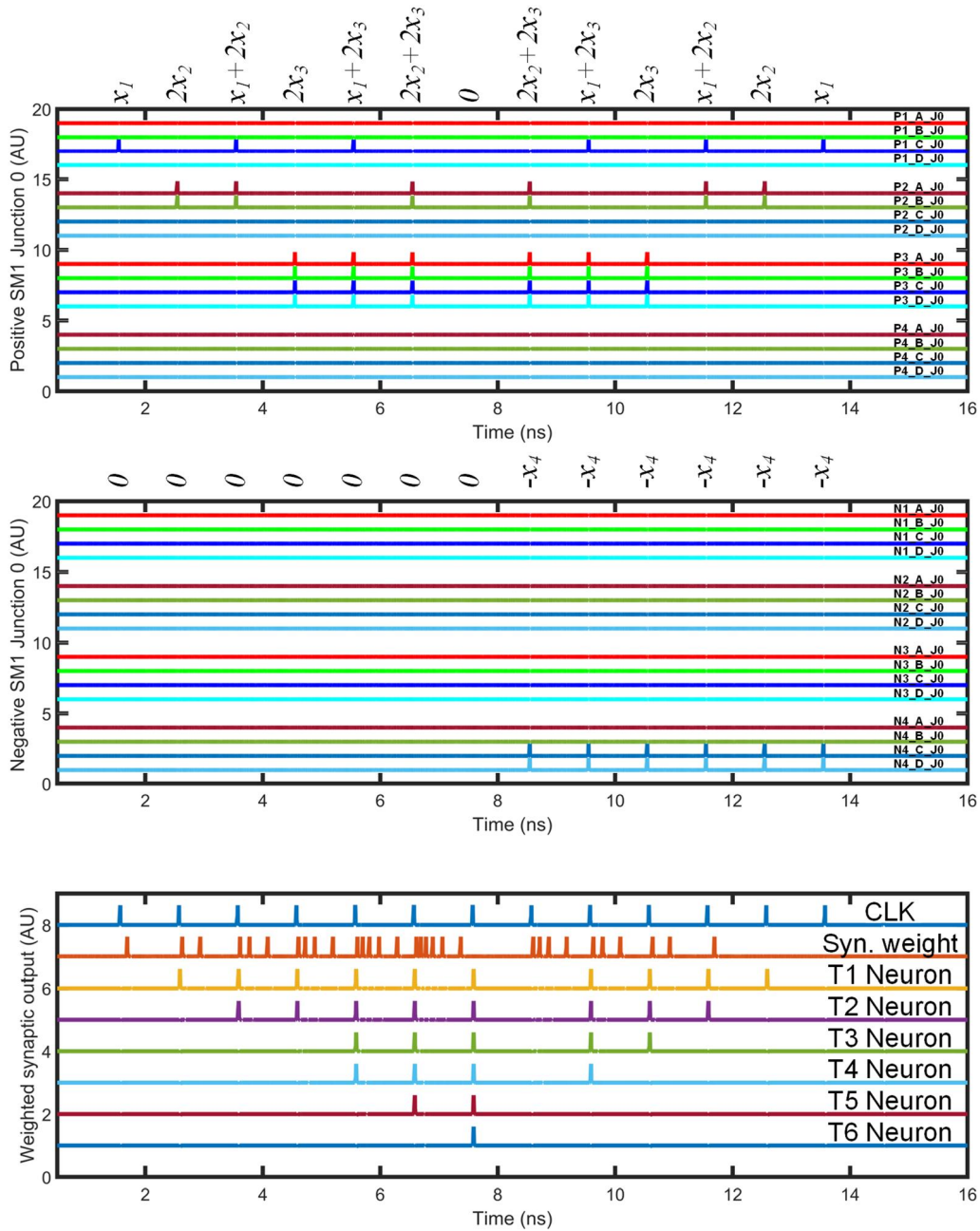

Figure 3 Simulation results for different weight and input combinations of a 4-input synapse (Figure 1) for $x_1=1$, $x_2=1$, $x_3=2$, $x_4=2$. Between 0 and 8ns, only excitatory weights are input (a). After 8ns, inhibitory weight of -1 is added (b). "Syn. Weight" line of (c) shows the output of the BQ circuit of Figure 1 which is the total synaptic weight of an input/weight combination. As shown, number of pulses are proportional to the total synaptic weight of each input/weight combination. T1-T6 Neuron lines show the outputs of 6 neurons with thresholds of 1 to 6. As shown, when the synaptic weight exceeds the threshold of the JJ-Soma, an output spike is fired. Clock frequency is set to 1 GHz. P1_A_J0 shows the J0 junction of the SM1-A cell of SM4-P1 in Figure 1.

## Axon and Multiple Connection from the Neurons

For SFQ pulse transmission, JTL circuits can be connected to JJ-Soma and JJ-Synapse cells and it will carry the roles of a biological neuron's axon part. If the distance to be transmitted is longer than a few hundreds of μm, passive transmission lines[32] may also be used. Splitter (SPL) circuits can be attached to the output sides of the JJ-Soma circuit to increase the fan-out to create the artificial axon ending structures.

## Artificial Neural Network Implementation

### Training

The training process in artificial neural networks refers to finding the optimum weight values all throughout the network such that the error between the actual outputs and the desired outputs for the provided input-output pairs are minimized, which is also called as "learning in neural networks". In general, Artificial Neural Networks are trained offline through an iterative weight update technique called backpropagation with gradient descent. Once the training error is reduced to an acceptable level during the learning process, the weight modification stops, and the resulting weight values are finalized [68].

We have adopted this training process for the optimum weight determination of our network. However, the weights obtained through backpropagation with gradient descent are real-valued and need to be discretized to be used with the Spiking Neural Network topology. This requires another round of optimization, in which the network weights are discretized. However, there is no standard technique to accomplish weight discretization that can guarantee the same network performance without any loss of accuracy. As a novel approach, we calculated the discrete weight values for our neural network through a traditional neural network training process followed by optimum weight discretization via genetic algorithms. After applying genetic algorithms, optimum discrete weights for our network became available as shown in Figure 2. Meanwhile, it is important to note that our optimum weight discretization algorithm for Spiking Neural Networks is not specific to our network; it can be used in other classification problems as well. The details of the weight determination process are explained in detail in Supplementary Information.

## Circuit Implementation

Now that we have a programmable spiking neuron and discretized training parameters, it is possible to implement the neural network circuit. From the training part, we have the neural network shown in Figure 2 with 4 inputs, 4 hidden layer neurons and 3 output neurons. Circuit schematics of the diagram is shown in Supplementary Information Figure S 7. Input values range from 0 to 2 and weight values range from -2 to 2. For the multiple connections from the hidden layer neurons to output neurons, SPL circuits are used. To set the weights, C1-C4 switches at each JJ-Synapse are programmed based on the tables in Figure 1(e,f). Hidden layer inputs may have values 0, 1, and 2 whereas the output layer may have input values 0 and 1. So, at the hidden layer SM4 based synapses and at the output layer, SM2 based synapses are used. As the soma part, we used programable JJ-MSoma with 1, 2, and 5 threshold values and selected the suitable one based on the network parameters.

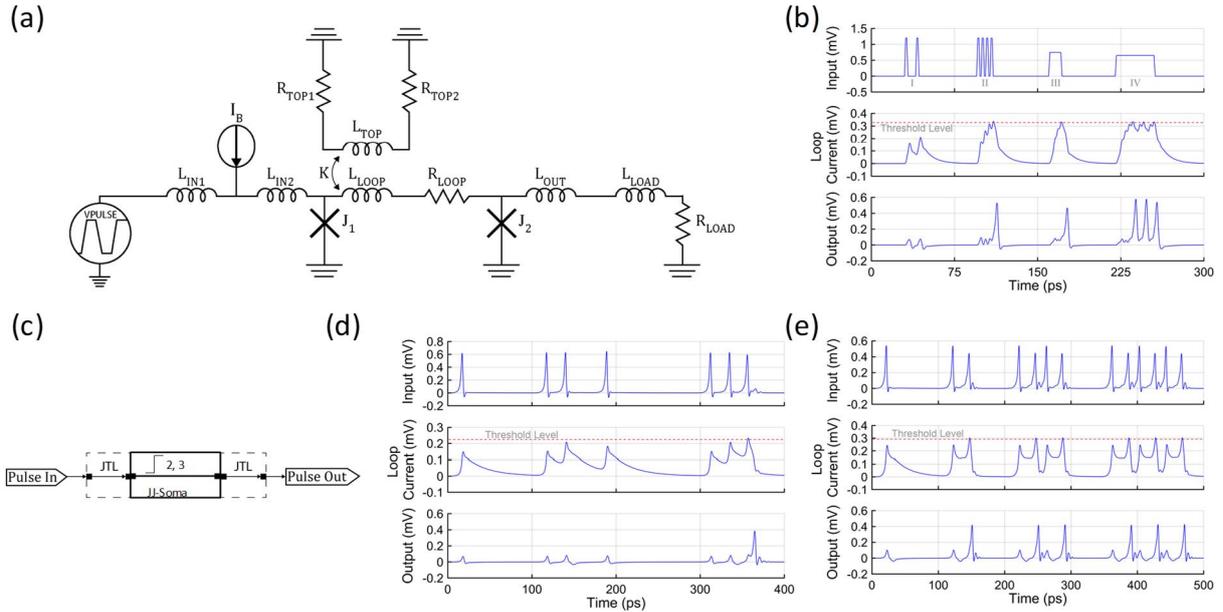

Figure 4 JJ-Soma Simulation Schematic (LLOAD=1 pH, LIN1=0.3 pH, LIN2=1.6 pH, LLOOP=9.8 pH, LOUT=1 pH, LTOP=9.6 pH, RLOAD=6.1 Ω, RLOOP=1.47 Ω, RTOP1=5 Ω, RTOP2=5 Ω, K=0.5 pH, J1= 243 µA, J2= 243 µA, Ib=0 µA) (a). JJ-Soma Simulation result with different input patterns (b). Test Schematic of JJ-Soma (c) (LIN1=0.3 pH, LIN2=1.11 pH, LLOOP =5.32 pH, LOUT =2.94 pH, LTOP=10.76 pH, RLOOP =0.34 Ω, RTOP1=0.31 Ω, RTOP2=0.3 Ω, K=0.21 pH, J1= 278 µA, J2= 272 µA, Ib=369 µA, Two pulse threshold) (LIN1=0.3 pH, LIN2=1.57 pH, LLOOP =9.42 pH, LOUT =4.59 pH, LTOP=12.34 pH, RLOOP =0.53 Ω, RTOP1=7.23 Ω, RTOP2=3.86 Ω, K=0.34 pH, J1= 150 µA, J2= 243 µA, Ib=342 µA, Three pulse threshold). Simulation results of the soma with 2-pulse threshold (c). The test for this soma has four different input situations: 1, 2, 4, and 6 pulses. As expected, after each two SFQ pulses, JJ-Soma fires an SFQ pulse. We set the interval time between two SFQ pulses in every pack of input to 20 ps. Due to this, JJ-Soma fires a single output every 40 ps and theoretical firing rate can be observed approximately 25 GHz. Simulation results of the soma with 3-pulse threshold (d). The test for this soma has three different input situations: 1, 3, and 3 pulses. As shown, when the number of pulses that arrive with proper timing is equal to the threshold of the JJ-Soma, it fires an SFQ pulse. Even if the number of pulses is equal to the threshold, the delay between pulses affects the result of soma output since the trapped current dissipates in the threshold loop. For the last state, we set the interval time between *SFQ pulses to 20ps.*

# Simulation Results
## JJ-Synapse

To verify the operation of a JJ-Synapse, the circuit shown in Figure 1-g is simulated and output of the buffer/quantizer (BQ) circuit is monitored for different weight and input combinations for a 4-input synapse as shown in Figure 3. Between 0 to 8ns, we considered only excitatory weights (Figure 3-a) that results in synaptic weights of 0 to 6. Then after 8ns we added an inhibitory weight of -1 (Figure 3-b) that resulted in total synaptic weights of 0 to 4 at the output. "Syn. Weight" line of Figure 3-c shows the output of the BQ circuit and the number of pulses are proportional with the synaptic weights. Then the output of the BQ circuit is connected to JJ-Soma circuits with thresholds from 1 to 6. As shown, when the synaptic weights exceed the thresholds of the JJ-Somas with 1 to 6 thresholds, an output spike is fired from the respective somas.

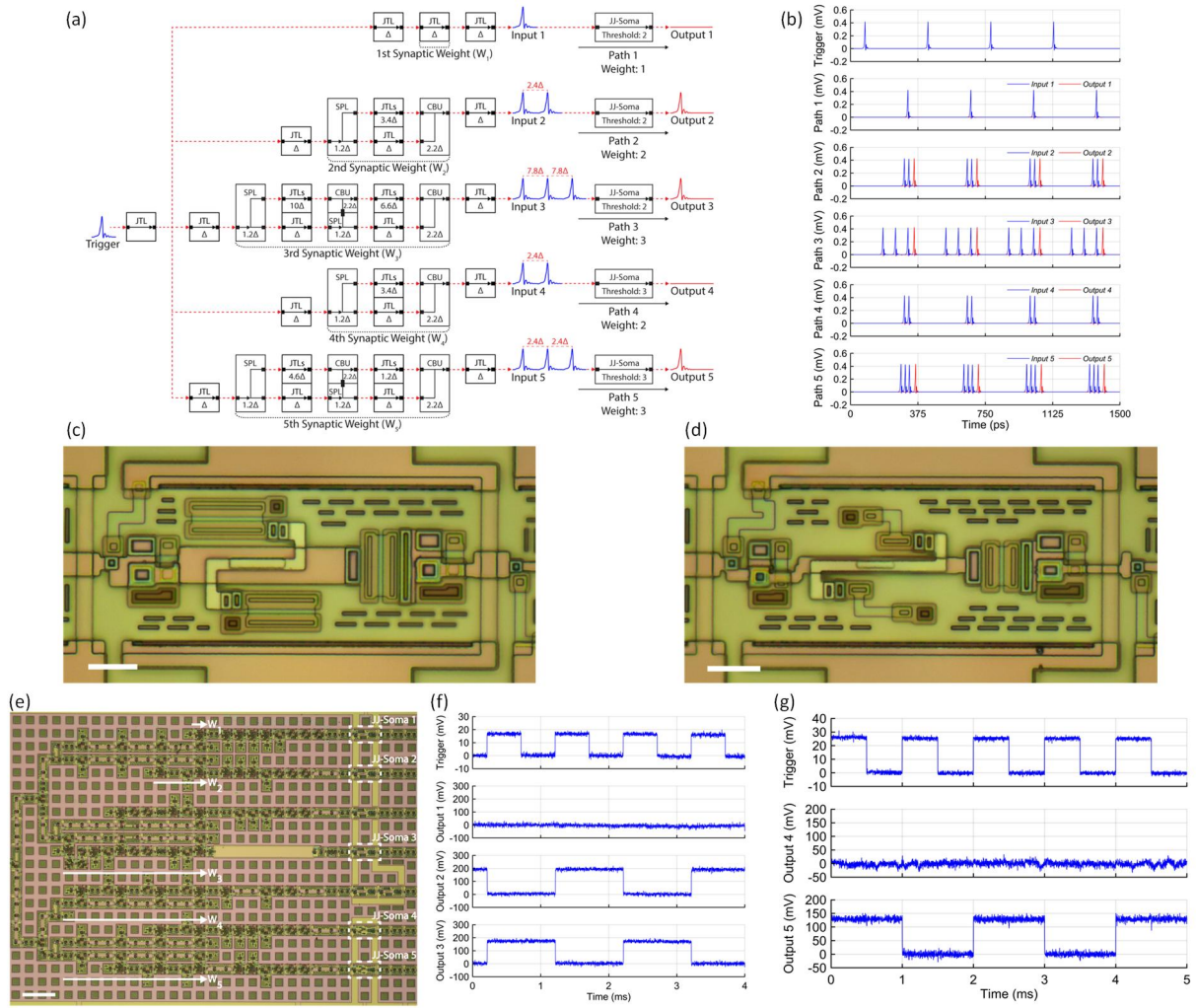

Figure 5 Individual soma test design (**a**). Simulation results of soma circuits with 2-pulse threshold and soma circuits with 3-pulse threshold with SFQ representation (**b**). Close-up view of 2-thresold JJ-Soma (Bar = 10 μm) (**c**), Close-up view of 3-thresold JJ-Soma (Bar = 10 μm) (**d**) JJ-Somas' test circuits (Bar = 100 μm) (**e**). Experimental results of soma circuits with two pulse threshold (**f**) and soma circuits with three pulse threshold (**g**) with DC representation.

## JJ-Soma

In JJ-Soma cell test is shown in Figure 4. Input pattern contains spikes and continuous signals to show the effect of different amplitudes, delays, and durations. First, two spikes with 1.2 mV amplitude arrive to the circuit (I). Due to insufficient number of spikes and large delay between spikes, JJ-Soma cell is unable to fire an output. However, if we increase the number of spikes to four and decrease the delay between spikes to 1 ps, neuron circuit can fire an output after reaching the threshold value (II). Likewise, the soma circuit can work under positive continuous signals. We put the soma circuit under test with the continuous inputs with different amplitudes and duration. First, we provided 0.75 mV continuous signal with 10 ps duration to observe a single output (III). Furthermore, we decreased the next continuous signal's voltage level to 65 mV and increased the total duration of signal to 30 ps (IV). For this situation, the circuit generated three SFQ pulses on output line. After 18 ps, the circuit generated the first SFQ pulse and created the next SFQ pulses every 10 ps. As a result, theoretical firing rate of the model without bias current was observed approximately 100 GHz in (IV) situation. Input and output voltage values can be observed on $V_{PULSE}$ and $J_2$. Additionally, JJ-Soma cell operation with inhibitory and excitatory inputs in the form of negative input voltages is given in the Supplementary Information Figure S *8*

Figure 4 shows two soma circuits with threshold of two and threshold of three SFQ pulses. For the first JJ-Soma circuit, after obtaining a single input pulse, second pulse should arrive within 65 ps. If it arrives after 65 ps, there will not be enough current in the threshold loop due to the decay of stored current and the circuit might require more SFQ pulses. For a soma of threshold two, after every two pulses, soma circuit provides a single pulse as an output and it is ready to obtain next pulse after releasing the output pulse. The soma of threshold three provides an output after obtaining three pulses and the delay between pulses should be set to maximum 20 ps.

SFQ test circuit schematics of the JJ-Soma together with peripheral circuits is shown in Figure 4(c) and simulation results of the soma circuits with two and three pulse thresholds are shown in Figure 4 (d, e). Input patterns that arrive to the soma are SFQ pulses from the previous JTL cell. With the arrival of an input pulse, circulating threshold loop current increases and stored current can be observed on LLOOP. If the current reaches the threshold limit of the loop, $J_2$ switches and fires an output pulse to next stage. For the experimental demonstration of the soma operation, the test circuit shown in Figure *5* (a) is designed. The pattern generator circuit generates 1, 2, and 3 pulses for 2 and 3 threshold soma circuits with an input trigger. Figure *5* (b) shows the input patterns arriving at each JJ-Soma and their outputs. If the number of pulses exceed the JJ-Soma threshold, then an SFQ pulse is released.

## Iris Network

After the implementation of JJ-Synapse and JJ-Soma, we designed the neural network circuit based on Figure 2 as shown in Supplementary Information Figure S 7. In the simulations, C1-C4 switches are modelled with resistors. When the switch is on, 1 Ω and when the switch is off, 1 MΩ values are assigned, respectively. For the test dataset of the IRIS dataset, we had 12 unique input cases. For each of them we simulated the circuit shown in Figure S 7 and obtained the results shown in Figure 6. All the output results were in line with the expected behavior, i.e. the triggered output neuron was correctly associated with the corresponding IRIS class in the training and test datasets.

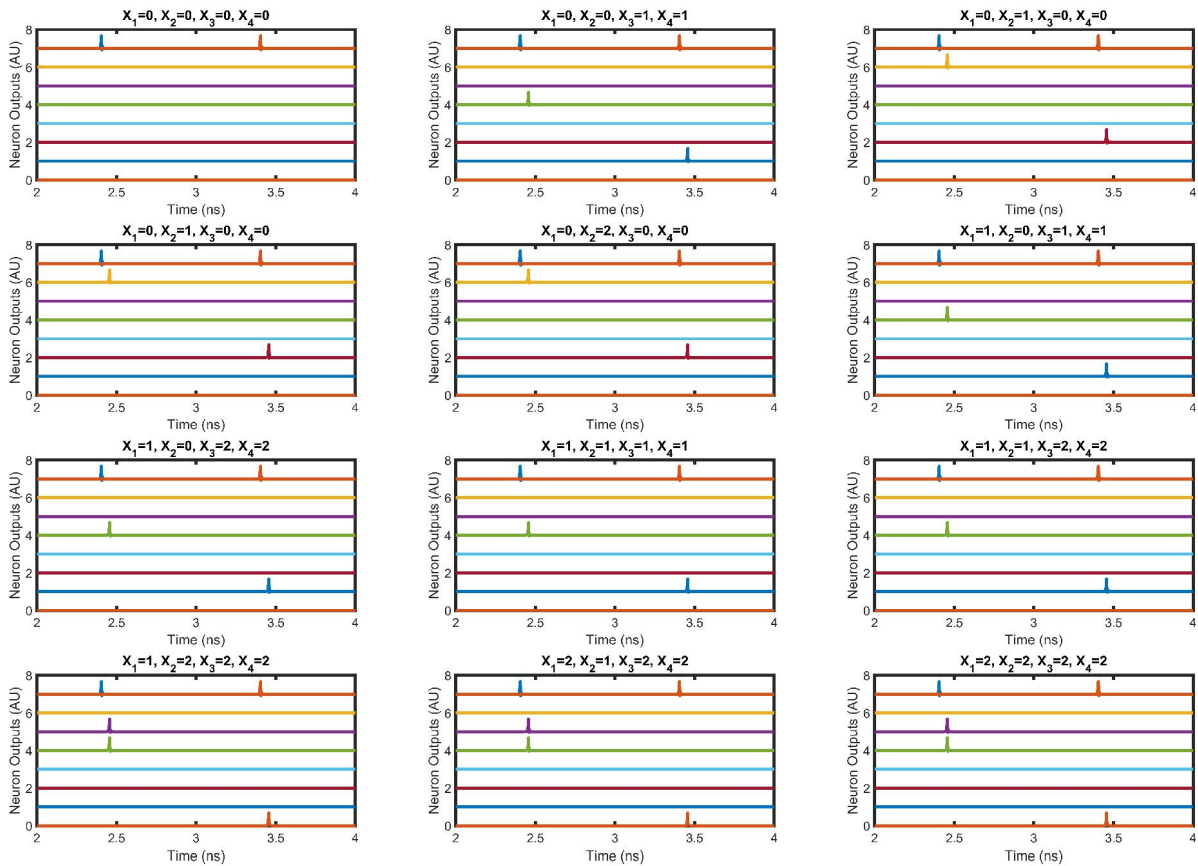

Figure 6 Simulation result of the test data of Iris dataset. The output (Y1, Y2, Y3) successfully determines the type of the Iris flower.

## Experimental Results

Each fabricated JJ-Soma cell covers 40 μm × 80 μm on-chip area. As explained in the previous section and shown in Figure *5*(a, b), a simple test structure that contains five JJ-Soma cells with different input patterns placed on the chip. Close-up view of the 2 and 3-thresold JJ-Somas are shown in Figure *5*(c, d). To trigger the cells with conventional signal generators, a DC-SFQ circuit and to measure the output of the soma circuits with an oscilloscope, an SFQ-DC circuit used at the input and output of neurons[29]. The details of these converter circuits are given in the Supplementary Information Figure S 9. With the trigger signal from the generator, the synapse circuits generated the pulse patterns and we observed outputs of JJ-Soma 1-5. Placement of the JJ-Soma in the chip is shown in Figure 5 (e). Experimental soma outputs which are consistent with the simulation results obtained for JJ-Soma circuit and the results are shown in Figure 5 (f, g). JJ-Soma 1 and 4 did not give any output as the input weights generated by the synapse circuit are lower than their thresholds and JJ-Soma 2, 3, and 5 gave output as the weights are equal or greater than the thresholds.

Note that, the trigger signal frequency set to 1 kHz in the experiments due to the limitations of the differential amplifier. However, this time scale is not representation of the firing rate of the neurons. In any case, regardless of the trigger signal frequency, the soma circuits must receive SFQ pulses up to 65 ps and 20 ps intervals from the synapses for 2-pulse and 3-pulse threshold soma circuits, respectively. In other words, a synaptic weight-2 means two SFQ pulses with maximum interval time of 65 ps and a synaptic weight-3 means three SFQ pulses with maximum interval time of 20 ps.

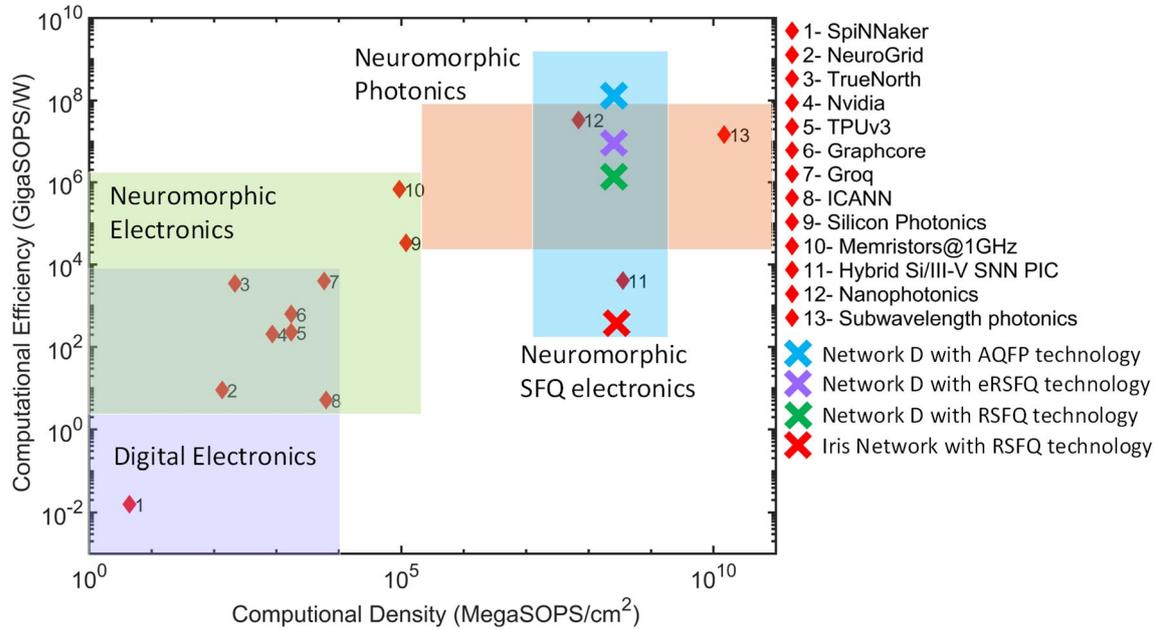

Figure 7 Benchmark of energy and speed of devices for neuromorphic applications. Adapted from[27]. For the power consumption of SFQ circuits, 400W/W cooling overhead is considered. (SpiNNaker[69–71], NeuroGrid[72,73], TrueNorth[74–76], Nvidia[77], TPUv3[78,79], Graphcore[80], Groq[81], HICANN[82,83], Silicon Photonics[84–86], Memristors@1GHz[87,88], Nanophotonics[89–91], Subwavelength photonics[92,93])

## Power consumption and scalability

For scalability and power consumption calculations, we considered two fully connected networks: one with 16 inputs, 2 hidden layers and 1 output layer with 16 neurons each (16×16×16 – Network A) and the other with 64 inputs with 3 hidden layers and 1 output layer with 64 neurons each (64×64×64×64 – Network B). Network A has 48 neurons and 4096 synapses whereas network B has 256 neurons and 16 million synapses. For fully connected design, neurons are terminated with 1 to 16 SPL and 1 to 64 SPLs respectively for network A and B.

Most of the power consumption in the circuit is the static power consumption due to the DC bias resistors. Bias currents and power consumptions of the individual cells are shown in Table S *3*. To calculate the power consumption of a neuron in Network A, for example, JJ-Soma, DFFe, JTL, BQ, and 1 to 16 SPL cells should be considered. SM1 cells have zero static power consumption as the Josephson junctions in the SQUID loop are in superconducting state when no input is applied. So, we need to calculate their dynamic power consumptions: During the generation of the SFQ pulse, JJ switches to voltage state and generates a quantum accurate digital signal in the form of single flux quanta $\Phi_0 = 2.068 \times 10^{-15}$ Wb. Energy per pulse is calculated by Eq. (3) where $I_C$ is the critical current of the JJs. Critical currents of the junctions in the SM1 circuits are 109 μA. So, energy per SFQ pulse is about $2.25 \times 10^{-19}$ Joule.

$$E_{SFQ} = \int_0^\tau I_c V dt = I_c \Phi_0 \tag{3}$$

In a hypothetical worst case, all the SM1 cells in either positive or negative side of the JJ-Synapse may switch. In the Iris network, there are 4×4×4 SM1 cells in the hidden layer, and 3×2×4 SM1 cells in the output layer. So, at each clock period 88 SFQ pulses may be generated. Then, total energy is 1.98×10$^{-17}$ Joule. With a 1 GHz clock rate, the dynamic power consumption is 20 nW. Similarly, for the case of 16×16×16 and 64×64×64×64 networks, dynamic power consumption is 0.12 μW and 0.57 mW, respectively.

Table S *4* shows the dynamic, static, and total power consumption of JJ-AN based neural networks with various complexities together with CMOS based implementations. We assumed 400W/W 4.2 K cooling overhead for the calculation of the total power consumption[94]. In this case, the total power consumptions are 14 mW, 158mW and 2000 mW respectively for Iris, NW-A and NW-B, respectively. Synaptic operations per second (SOPS) is calculated as the number of switching events at the synapses per second. For 1 GHz clock rate, SOPS are 1.2×10$^{10}$, 4×10$^{12}$, and 1.6×10$^{16}$ respectively for these three networks. Then, energy efficiencies are 8.57×10$^{11}$, 2.53×10$^{13}$, and 8×10$^{15}$ SOPS/W including the cooling overhead respectively.

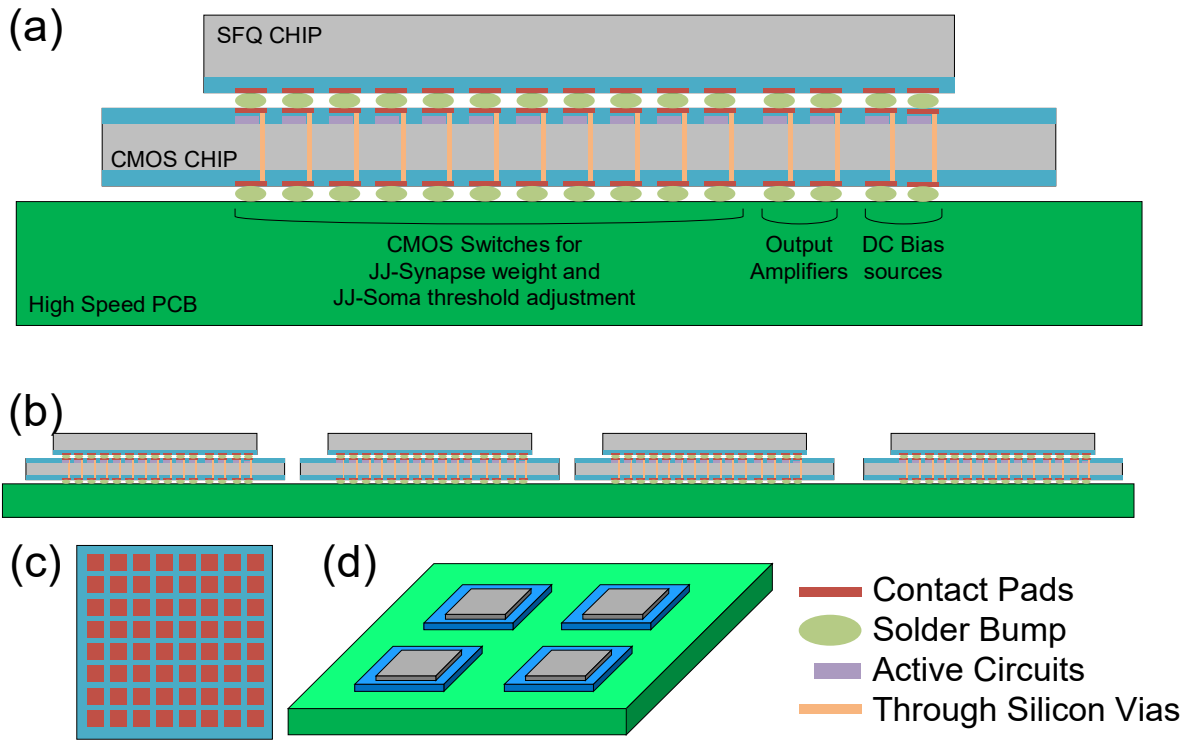

Figure 8 Multichip module packaging concept. Single core concept (a). Multicore concept (b). Top and 3D view (c,d).

We have also estimated the scale of the network that can be implemented for a 2W off-the shelf cryocooler. In this case, it is possible to implement a 256-core network with 256 neurons each, which provides about 10$^{18}$ SOPS computation power and 10$^{15}$ SOPS/W energy efficiency (Network D). If this circuit was to be implemented by using energy efficient RSFQ (eRSFQ)[95] or adiabatic quantum flux parametron (AQFP)[96] technologies, then the estimated energy efficiencies would be 10$^{16}$ SOPS/W and 10$^{17}$ SOPS/W respectively. (See Supplementary information for calculation details). Figure 7 which is adapted from [27] with the addition of neuromorphic SFQ electronics, shows the benchmark of energy and speed of devices for neuromorphic applications. As shown, SFQ electronics is one of the most promising technologies for implementing high performance neuromorphic processors.

To realize the training and programming of the neural network, we propose to use a multi-chip module (MCM) configuration with cryo-CMOS circuits as shown in Figure 8. Switching speed of 1 GHz is easily achievable with cryo-CMOS circuits. Successful demonstrations of MCMs up to 15 μm pitch has already been shown[33]. For Network B, we need 131,072 switches to control the synaptic weights in a 362×362 contact pad configuration. If we assume a pitch size of 50 μm for the solder bumps, then we need about 18mm×18mm chip space for contacts. Similarly, we need some contact pads for DC bias sources (about two for each neuron) for the circuits and output amplifiers (one for each output neuron). Roughly, we will need an array of 400×400 for the contacts as shown in Figure 8(a). With a pitch size of 50 μm, it makes a 2cm×2cm chip area which is a reasonable dimension. CMOS chip may have through silicon vias (TSV) for connection from bottom side to a master package as shown in Figure

8(b). If we consider this module as one core of a neuromorphic processor, than SFQ/CMOS hybrid module can be stacked on the master package as shown in Figure 8(c, d) for higher complexity neural networks.

## Discussion

Proposed JJ-Neuron has all the fundamental requirements of an artificial neuron: *i)* adjustable threshold *ii)* ability of synaptic weighting *iii)* fan-out to many neurons, *iv)* ability to handle various input levels *v)* ability to train the network at run time. In addition, it is possible to design the neural network by using established neural network design methods in a transparent manner. For the control of the network, simple CMOS switches are sufficient, and performance/dimensional requirements are within the practical limits of a CMOS circuits and flip-chip technology. Both the SFQ and CMOS circuits are compatible with multi-project-wafer services that enable convenient, low cost and mass production of the chips.

When the circuit is scaled to operate within the limits an off-the-shelf cryocooler with 2 W excess cooling power, it has the potential to have state of the art SOPS and SOPS/W performance levels of $10^{18}$ and $10^{15}$ SOPS/W respectively even with the cooling overhead. Furthermore, the proposed topology is quite suitable for e-RSFQ and AQFP type implementations since the operating frequency is about 1 GHz. In this case, it is possible to have zero static power consumption and increase the power efficiency to $10^{16}$ SOPS/W and $10^{17}$ SOPS/W respectively.

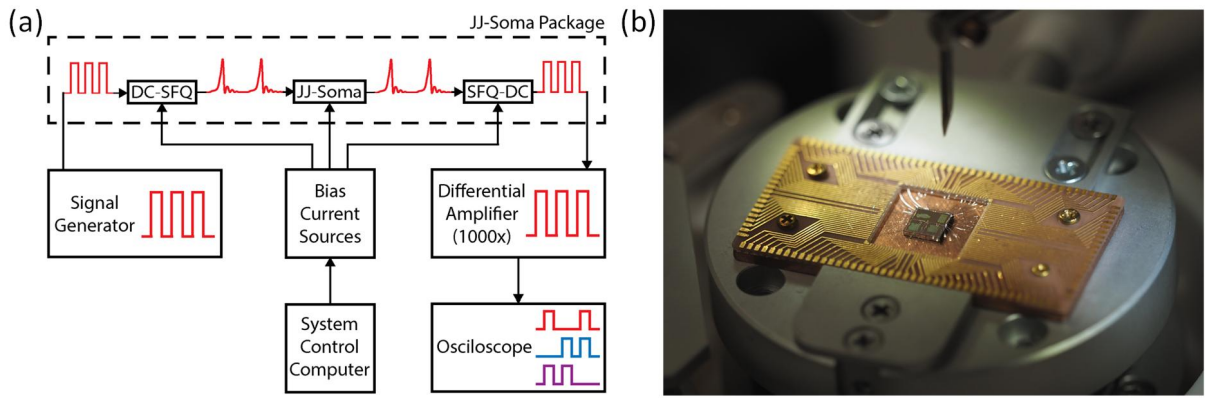

Figure 9 Block diagram of the test system (**a**). JJ-Soma package photo (**b**).

## Methods

### Simulations

All simulations are carried out by using Josephson Simulator (JSIM)[97] and we used commercially available numerical computing environment for user interface of tools and optimization of the circuits.

### Neural Network Design

The mathematical (software) representation of the neural network including weight determination and optimum discretization is accomplished with MATLAB software. First, the 4x4x3 traditional neural network is initialized with random weights. Then, backpropagation learning is adopted to determine the real-valued weight values for the network. Finally, genetic algorithm is implemented to find the optimum quantization levels for each neuron so that each weight value within the network becomes one of -2, -1, 0, 1, 2, the discrete weight values selected for this particular problem. The details about the weight discretization and optimization through genetic algorithms is given in Supplementary Information.

### Fabrication

The circuits were fabricated in the clean room for analog-digital superconductivity (CRAVITY) of National Institute of Advanced Industrial Science and Technology (AIST) with the standard process 2 (STP2)[53]. The AIST-STP2 is based on the Nb circuit fabrication process developed in International Superconductivity Technology Center (ISTEC).

### Measurements

In this research, all the experiments were completed in a two-stage pulse tube cryocooler (Sumitomo RP-062B) with a temperature stability of about 10 mK and excess cooling power of about 100 mW at a temperature of 4.2 K. Coaxial RF cables created input and output signal connections between the room temperature electronics and chip. As DC bias lines, Phosphor-Bronze wires were placed. The equipment always stays in the Faraday cage. A three-layer μ-metal magnetic shield around the chip was used to attenuate the external magnetic field to about 5 nT. The block diagram and JJ-Soma package photo are given in Figure 9 and details of the experimental setup are explained in [98].

## Acknowledgements
Authors would like to thank Eren Can Aydogan for providing the counter, Yigit Tukel for developing the original PSO algorithm for SFQ circuits used in this study. We are grateful to Dr. Mehmet Unlu for critically reading the manuscript.

## Author contributions statement
AB and MAK designed and performed the experiments for the JJ-Soma circuits. AB and SR designed JJ-Synapse and Iris network circuits. MO designed the SNN for Iris dataset.

## Additional information
**Competing interests:** The authors have no conflicts of interest to disclose. TOBB University of Economics and Technology has filed the following patent applications
  i) TR patent application no. 2018/09195, 28 June 2018, PCT patent application no. PCT/TR2019/050510 with A.B. and M.A.K. as the named inventors based on this work.
  ii) TR patent application no. 2019/10971, 22 July 2019, A.B. and M.A.K. as the named inventors based on this work.
  iii) TR patent application no. 2020/08489, 02 June 2020, with A.B. and SR as the named inventors based on this work.
  iv) TR patent application no. 2020/10176, 29 June 2020, with A.B. and SR as the named inventors based on this work.

**Data and materials availability:** All data needed to evaluate the conclusions are present and additional data related to this paper can be requested from the authors. IRIS dataset from UCI Machine Learning Repository is available from http://archive.ics.uci.edu/ml/datasets/iris. Raw data for the simulation and experiments are available at doi:10.6084/m9.figshare.12599756.

**Code availability:** Opensource software, Jsim[99] is used for the circuit simulations. Commercial software, Matlab 2018b with Deep Learning Toolbox and Global Optimization Toolbox, is used for the weight discretization and neural network design. The MATLAB code used to train the neural network based on the explanations of the Supplementary Information is available at doi: 10.6084/m9.figshare.12599777.